\documentclass{article}
\usepackage[T1]{fontenc}
\usepackage{lmodern}
\usepackage{spconf,cite}
\usepackage{amsmath,amssymb,amsfonts}
\usepackage{algorithmic}
\usepackage{graphicx}
\usepackage{textcomp}
\usepackage{xcolor}
\usepackage{bm}
\usepackage{siunitx}
\robustify\bfseries
\usepackage{booktabs}
\usepackage{multirow}
\usepackage{makecell}
\usepackage{pifont}
\usepackage{threeparttable}
\usepackage{balance}
\usepackage[capitalize,nameinlink]{cleveref}
\usepackage[ruled,vlined,linesnumbered]{algorithm2e}
\usepackage{listings}
\usepackage[dvipsnames]{xcolor}
\crefname{equation}{}{}
\Crefname{equation}{Equation}{Equations}
\usepackage{tikz}
\usetikzlibrary{arrows.meta,calc,positioning,shapes.geometric}
\usepackage[table]{xcolor}

\definecolor{ai2offwhite}{HTML}{FBF4EE}

\newcommand*{\metric}[2]{%
  \fpeval{round(#1,2)} \pm \fpeval{round(#2,2)}%
}
\usepackage{IEEEtrantools}

\let\oldthebibliography\thebibliography
\renewcommand{\thebibliography}[1]{%
  \oldthebibliography{#1}%
  \vspace{-0.5em}%
  \setlength{\itemsep}{0.2pt}%
  \setlength{\parskip}{0.2pt}%
}

\usepackage{etoolbox}
\makeatletter
\patchcmd{\@makecaption}
  {\vskip 10pt}
  {\ifdefstring{\@captype}{table}{\vskip 3.8pt}{\vskip 10pt}}
  {}
  {\PackageWarning{captionpatch}{Failed to patch \string\@makecaption}}
\makeatother

\title{Noisy Target Training Meets Drifting\\for Weakly-Supervised Speech Enhancement}
\name{{\shortstack[c]{
    Yoshiki Masuyama$^{1}$,
    Dominik Klement$^{1,2}$%
    \sthanks{This work was done during an internship at MERL.},
    Christoph Boeddeker$^{1}$,
    Julius Richter$^{1}$, \\
    Kohei Saijo$^{3}$,
    Gordon Wichern$^{1}$,
    Takahiro Edo$^{1,3}$,
    Jonathan Le Roux$^{1}$
}}}
\address{$^{1}$Mitsubishi Electric Research Laboratories (MERL), USA \\
$^{2}$Speech@FIT, Brno University of Technology, Czechia, \quad
$^{3}$Mitsubishi Electric Corporation, Japan
}
\begin{document}
\ninept
\setlength{\abovedisplayskip}{5pt}
\setlength{\belowdisplayskip}{5pt}
\bstctlcite{BSTcontrol}
\maketitle
\begin{abstract}
In speech enhancement, standard supervised training requires pairs of noisy and clean speech. Such pairs are typically created by artificially degrading clean speech, leading to a domain mismatch between synthetic training data and real recordings. Noisy target training (NyTT) mitigates this limitation by training an enhancement model to predict a noisy recording from a noisier version generated by adding extra noise. However, since the training target is noisy, NyTT might learn to leave residual noise in the enhanced output. In this paper, we address this issue in NyTT by incorporating a Sinkhorn drifting loss that matches the distribution of the enhanced speech to that of unpaired clean speech. The Sinkhorn drifting loss attracts enhanced speech toward high-density regions of the clean-speech distribution while repelling it from the current enhanced-speech distribution in latent space. This loss uses only unpaired clean speech rather than the clean counterpart of each degraded recording, preserving the weakly-supervised nature of NyTT. Experimental results on VoiceBank-DEMAND and CHiME-3 real recordings demonstrate the efficacy of the proposed combination.
\end{abstract}
%
\section{Introduction}
\vskip -1mm

Speech enhancement aims to suppress background noise and improve the quality of degraded speech recordings~\cite{Gannot2017,Wang2018review}.
Modern speech enhancement models are typically trained in a supervised manner, which requires pairs of degraded speech and underlying clean speech.
We refer to this standard supervised training
strategy as clean-target training (CTT), following \cite{fujimura2021nytt}.
Since the underlying clean speech is inaccessible in real environments, CTT relies on synthetic pairs, where clean speech is artificially degraded, e.g., by adding a separate noise recording.
Consequently, supervised models might suffer from performance degradation on in-the-wild recordings due to mismatches in acoustic conditions~\cite{leglaive2023chime7t2}.

To leverage real-world noisy speech recordings, various weakly-supervised and unsupervised training paradigms have been developed~\cite{fujimura2021nytt,tzinis2022remixit,jiang2023unse,klement2025}.
Noisy target training (NyTT) is simple yet shows promising performance~\cite{fujimura2021nytt}.
NyTT trains an enhancement model to predict an observed signal from a noisier version generated by adding extra noise.
Since the training target is noisy, the enhanced speech tends to retain residual noise, limiting perceptual quality~\cite{maciejewski2026,jiang2026}.
Although NyTT variants have mitigated this issue using mean-teacher techniques~\cite{fujimura2025nyttext,jiang2026,song2026artt}, their objectives still do not explicitly encourage the distributions of clean and enhanced speech to match.

\begin{figure}[t!]
\centering
\input{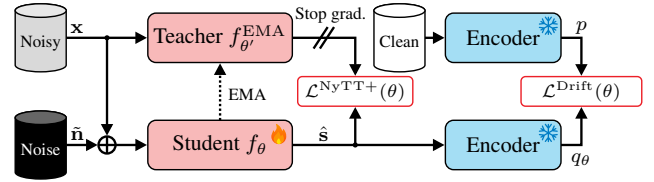}
\vskip -1mm
\caption{
Overview of the proposed NyTT with unpaired drifting.
}
\label{fig:overview}
\vskip -2mm
\end{figure}

Generative speech enhancement~\cite{lemercier2025diffusion} has gained increasing attention and can produce perceptually convincing speech even with one or a few inference steps~\cite{lay2024single,wang2026meanse}.
In particular, DriftSE~\cite{xu2026speech,xu2026driftsejournal} demonstrates promising one-step performance based on drifting models~\cite{deng2026generative}.
Drifting models define a vector field, called a drifting field, that encourages the model-induced enhanced-speech distribution $q_{\bm{\theta}}$ to move toward the clean-speech distribution $p$.
In addition to its strong performance in the CTT setting, DriftSE also shows potential for training an enhancement model using observed noisy signals and unpaired clean speech.
On VoiceBank-DEMAND (VBDM)~\cite{valentinibotinhao2016}, it achieves a DNSMOS P.808~\cite{reddy2021dnsmosp808} score of 3.61~\cite{xu2026speech}, indicating improved perceptual quality.
However, SI-SDR and WER are degraded even relative to the original noisy signal~\cite{xu2026driftsejournal}.
These results suggest that unpaired drifting alone is insufficient to preserve speech content, despite its ability to improve the perceptual quality.

In this paper, we propose incorporating a drifting loss into NyTT to suppress residual noise and improve the perceptual quality of enhanced speech, as illustrated in \cref{fig:overview}.
The proposed method builds on Sinkhorn drifting~\cite{he2026sinkhorn}, a variant of drifting models in which the attraction--repulsion field is induced by the Sinkhorn divergence~\cite{feydy2019interpolating}.
The drifting loss allows us to match the distribution of the enhanced speech to that of unpaired clean speech without requiring clean counterparts for the noisy training data.
We expect the waveform-level weak supervision provided by NyTT and the distribution-level supervision provided by Sinkhorn drifting to complement each other.
In addition, we propose to compute the drifting field in the latent space of a speech quality assessment (SQA) model, where distances are expected to reflect perceptual naturalness.
We validate our approach on the synthetic mixtures from VBDM and demonstrate its efficacy on real recordings from CHiME-3.

\section{Related Work}

\subsection{Noisy target training (NyTT) and its variants}

We model an observed signal $\mathbf{x}\!\sim\!p_\text{noisy}$ as the sum of underlying clean speech $\mathbf{s}\!\in\!\mathbb{R}^L$ and background noise $\mathbf{n}\!\in\!\mathbb{R}^L$,
where $L$ denotes the length of the signal.
Standard supervised training for speech enhancement, also referred to as CTT, optimizes an enhancement model $f_{\bm{\theta}}$ with parameters $\bm{\theta}$ to minimize the prediction error with respect to the underlying clean speech:
\begin{equation}
    \mathcal{L}^\mathrm{CTT}(\bm{\theta}) = \mathbb{E}_{(\mathbf{x}, \mathbf{s})} \mathcal{D} (\mathbf{s}, f_{\bm{\theta}} (\mathbf{x})),
    \label{eq:ct}
\end{equation}
where $\mathcal{D}$ denotes a function measuring the prediction error, e.g., the negative signal-to-noise ratio (SNR).
This training process requires pairs of $(\mathbf{x}, \mathbf{s})$, but the underlying clean speech $\mathbf{s}$ is not accessible in real noisy environments.
Hence, the pairs are artificially generated by adding noise $\mathbf{n}\!\sim\!p_\text{noise}$ to pre-recorded clean speech $\mathbf{s}\!\sim\!p$.
However, the domain mismatch with real-world acoustic conditions may lead to performance degradation on
real recordings~\cite{leglaive2023chime7t2}.

NyTT relaxes the requirement of CTT by utilizing in-domain noisy speech as the training target instead of the clean speech~\cite{fujimura2021nytt}:
\begin{equation}
    \mathcal{L}^\mathrm{NyTT}(\bm{\theta}) = \mathbb{E}_{\mathbf{x}, \tilde{\mathbf{n}}} \mathcal{D} (\mathbf{x}, f_{\bm{\theta}} (\mathbf{x} + \tilde{\mathbf{n}})),
    \label{eq:nytt}
\end{equation}
where $\tilde{\mathbf{n}}\!\sim\!p_\text{noise}$ is an additional noise signal for generating a noisier model input $\mathbf{x} + \tilde{\mathbf{n}}$.
The model is expected to suppress both ${\mathbf{n}}$ and $\tilde{\mathbf{n}}$ as they are not easy to separate when both are from the same distribution.
However, since the target is noisy, NyTT often learns to leave residual noise in the enhanced output.

Recent studies theoretically show that the optimal enhanced signal for NyTT leaves a certain amount of observation noise and additional noise~\cite{jiang2026,maciejewski2026}.
To mitigate this issue, NyTT variants adopt iterative refinement~\cite{fujimura2025nyttext,jiang2026} and a mean-teacher framework~\cite{song2026artt}.
In the mean-teacher framework, the teacher parameters $\bm{\theta}'$ are updated based on student parameters $\bm{\theta}$ using an exponential moving average (EMA) after each optimization step.
The training target for NyTT is replaced by the output of the teacher model for the observed signal:
\begin{equation}
    \mathcal{L}^\mathrm{NyTT+}(\bm{\theta}) = \mathbb{E}_{\mathbf{x}, \tilde{\mathbf{n}}} \mathcal{D} (\texttt{sg}(f_{\bm{\theta}'}^\mathrm{EMA}(\mathbf{x})), f_{\bm{\theta}} (\mathbf{x} + \tilde{\mathbf{n}})),
    \label{eq:nyttsd}
\end{equation}
where $\texttt{sg}$ denotes the stop-gradient operation.
The teacher network suppresses the observation noise $\mathbf{n}$ in $\mathbf{x}$, providing a clearer pseudo target~\cite{fujimura2025nyttext,jiang2026,song2026artt}.
These NyTT variants have improved enhancement performance further, but these methods do not explicitly encourage the enhanced speech to follow the clean-speech distribution.

\subsection{Speech enhancement training with unpaired clean speech}

Another direction for weakly-supervised speech enhancement exploits unpaired clean speech~\cite{jiang2023unse,klement2025} and aligns the enhanced-speech distribution with the clean-speech one.
For example, unSE~\cite{jiang2023unse} uses adversarial training, and its enhancement-model loss contains a regularization term for reconstructing the input itself as follows:
\begin{equation}
    \mathcal{L}^\mathrm{unSE}(\bm{\theta}) = \mathbb{E}_{\mathbf{x}} \left[ \mathcal{D} (\mathbf{x}, f_{\bm{\theta}} (\mathbf{x})) - d_{\bm{\phi}}(f_{\bm{\theta}} (\mathbf{x})) \right],
    \label{eq:unse}
\end{equation}
where $d_{\bm{\phi}}$ is a discriminator with parameters $\bm{\phi}$ optimized together with the enhancement model in an adversarial way.
While adversarial training aims to match the two distributions, the data-fidelity term constrains the output to remain close to the noisy observation $\mathbf{x}$, which can hinder the denoising capability.
Recent work~\cite{klement2025} partially mitigates this issue by simultaneously predicting the noise, but the entire training becomes more complicated with multiple discriminators.
These methods aim to learn the clean-speech distribution from unpaired data but rely on sensitive adversarial training.

\section{Methods}

\subsection{Speech enhancement based on drifting models}

DriftSE~\cite{xu2026speech} formulates speech enhancement as matching the distribution of enhanced speech $q_{\bm{\theta}}$ to the clean-speech distribution $p$, using drifting models~\cite{deng2026generative} instead of adversarial training.
Here, $q_{\bm{\theta}}$ is the pushforward distribution~\cite{villani2009optimal} induced by applying the enhancement model $f_{\bm{\theta}}$ to noisy signals distributed according to $p_\text{noisy}$:
\begin{equation}
q_{\bm{\theta}} = (f_{\bm{\theta}})_\# p_\text{noisy}.
\end{equation}
That is, for a noisy input signal $\mathbf{x}\!\sim\!p_\text{noisy}$, the corresponding enhanced signal $f_{\bm{\theta}}(\mathbf{x})$ is distributed according to $q_{\bm{\theta}}$.

Then, drifting models introduce a drifting field $\mathbf{V}_{ q_{\bm{\theta}}, p}: \mathbb{R}^D \rightarrow \mathbb{R}^D$ in the latent space of a pretrained encoder $h$ with feature dimension $D$.
The drifting field is designed to satisfy the following equilibrium condition~\cite{deng2026generative}:
\begin{equation}
    q_{\bm{\theta}} = p \quad \Longrightarrow \quad \mathbf{V}_{q_{\bm{\theta}}, p}(\mathbf{z})=0,
\quad \forall \mathbf{z} \in \mathbb{R}^{D}.
\label{eq:equilibrium}
\end{equation}
Hence, when $q_{\bm{\theta}}$ matches $p$, the drifting field vanishes everywhere, reaching an equilibrium point.

In the original drifting framework~\cite{deng2026generative}, the drifting field $\mathbf{V}_{q_{\bm{\theta}}, p}$ is derived from the classical mean-shift method~\cite{fukunaga1975estimation}.
Specifically, it is formulated as a combination of an attractive term that pulls samples toward the target distribution $p$ and a self-correction term that accounts for the current model distribution $q_{\bm{\theta}}$~\cite{deng2026generative}:
\begin{equation}
\mathbf{V}_{q_{\bm{\theta}}, p}(\mathbf z)
=
\mathbf V_p(\mathbf z)
-
\mathbf V_{q_{\bm{\theta}}}(\mathbf z),
\label{eq:orignial_drifting_field}
\end{equation}
where $\mathbf V_\rho(\mathbf z)$ for $\rho \in \{p,q_{\bm{\theta}}\}$ is given by
\begin{equation}
\mathbf V_\rho(\mathbf z)
=
\frac{
\mathbb E_{\tilde{\mathbf  z}\sim \rho}
\!\left[
\tilde{\mathbf z}\,k(\tilde{\mathbf z} - \mathbf z)
\right]
}{
\mathbb E_{\tilde{\mathbf z}\sim \rho}
\!\left[
k(\tilde{\mathbf z} - \mathbf z)
\right]
}
-
\mathbf z,
\label{eq:mean_shift_field}
\end{equation}
and $k$ is a similarity kernel.

In DriftSE~\cite{xu2026speech}, the speech enhancement model $f_{\bm{\theta}}$ is then trained to minimize the following loss function:
\begin{equation}
    \mathcal{L}^\mathrm{DriftSE}(\bm{\theta}) = \mathbb{E}_{(\mathbf{x}, \mathbf{s}), \bm{\epsilon}} \left\| \hat{\mathbf{z}} - \texttt{sg}(\hat{\mathbf{z}} + \mathbf{V}_{ q_{\bm{\theta}}, p}(\hat{\mathbf{z}}) ) \right\|_2^2,
\label{eq:original_loss}    
\end{equation}
where the clean speech samples $\mathbf{s}$ are used for computing the drifting field, $\hat{\mathbf{z}}$ is an enhanced-speech feature extracted by a pre-trained speech encoder $h$ (e.g., DistillHuBERT~\cite{chang2022}):
\begin{equation}
    \hat{\mathbf{z}} = h(f_{\bm{\theta}}(\mathbf{x} + \bm{\epsilon})),
    \label{eq:encoding}
\end{equation}
and $\bm{\epsilon}\!\in\!\mathbb{R}^L$ is Gaussian noise to introduce stochasticity.
This training objective can be viewed as updating the model parameters $\bm{\theta}$ so that the enhanced speech samples move toward clean-speech distribution.
In the unpaired setting, the clean speech samples $\mathbf{s}$ in \eqref{eq:original_loss} are sampled independently of $\mathbf{x}$, potentially from a different dataset~\cite{xu2026speech}.

\subsection{Drifting field induced by Sinkhorn divergence}
\label{sec:sinkhorn_drifting}

We adopt a variant of the drifting framework called Sinkhorn drifting~\cite{he2026sinkhorn}.
It uses a loss function similar to the original drifting loss in \eqref{eq:original_loss} but derives the drifting field from the gradient flows induced by the Sinkhorn divergence~\cite{feydy2019interpolating}.
This additionally ensures that zero drift implies an exact match between the distributions, i.e., the converse of \cref{eq:equilibrium}, under mild assumptions~\cite{he2026sinkhorn}.
In addition to this theoretical guarantee, Sinkhorn drifting improves the training stability~\cite{gretton2026wasserstein}.

The debiased Sinkhorn divergence between $p$ and $q_{\bm{\theta}}$ is defined using the entropically regularized optimal transport cost $\mathrm{OT}_\tau$~\cite{feydy2019interpolating}:
\begin{equation}
\!\!S_\tau(q_{\bm{\theta}},p)
=
\mathrm{OT}_\tau(q_{\bm{\theta}},p)
-\frac{1}{2}\mathrm{OT}_\tau(q_{\bm{\theta}},q_{\bm{\theta}})
-\frac{1}{2}\mathrm{OT}_\tau(p,p),
\label{eq:sinkhorn_divergence_model}
\end{equation}
where $\tau>0$ controls the strength of the entropic regularization~\cite{peyre2019computational}.
Intuitively, Sinkhorn drifting defines the drifting field using the gradient flow of \cref{eq:sinkhorn_divergence_model}, and then the model parameters $\bm{\theta}$ are updated so that the model-induced enhanced-speech distribution approximates one forward Euler step of the gradient flow.

In practice, we need to approximate the distributions $q_{\bm{\theta}}$ and $p$ by mini-batch empirical measures. 
Hence, given a mini-batch of size $B$, the two distributions are approximated by
\begin{equation}
\hat q_{\bm{\theta}}
=
\frac{1}{B}\sum_{i=1}^{B}\delta_{\hat{\mathbf z}^{(i)}}, \quad
\hat p
=
\frac{1}{B}\sum_{j=1}^{B}\delta_{\mathbf z^{(j)}},
\end{equation}
where $\delta_\mathbf{z}$ denotes the Dirac measure, $\hat{\mathbf z}^{(i)}$ denotes an enhanced-speech feature, and $\mathbf z^{(j)}$ denotes a feature from clean speech. 
Then, the velocity field $\mathbf{V}_{q_{\bm{\theta}}, p}(\mathbf z)$ is formulated by using the optimal couplings for the entropically regularized optimal transport problems in \eqref{eq:sinkhorn_divergence_model} with the quadratic cost $C(\mathbf z,\mathbf z')=1/2\,\| \mathbf z-\mathbf z'\|_2^2$.
Specifically, the optimal couplings are approximated by using Sinkhorn algorithm with $\ell$ iterations followed by a row-wise scaling, resulting in $\pi_{\hat q_{\bm{\theta}},\hat p}^{\ell}$ and $\pi_{\hat q_{\bm{\theta}},\hat q_{\bm{\theta}}}^{\ell}$.
The Sinkhorn drifting field is obtained as
\begin{multline}
\mathbf V_{\hat q_{\bm{\theta}},\hat p}^{\ell}(\hat{\mathbf z}^{(i)})
=
\sum_{j=1}^{B}
(B\pi_{\hat q_{\bm{\theta}},\hat p}^{\ell})_{ij}
\left({\mathbf z}^{(j)}-\hat{\mathbf z}^{(i)}\right) \\
-
\sum_{j=1}^{B}
(B\pi_{\hat q_{\bm{\theta}},\hat q_{\bm{\theta}}}^{\ell})_{ij}
\left(\hat{\mathbf z}^{(j)}-\hat{\mathbf z}^{(i)}\right).
\label{eq:sinkhorn_drift_field}
\end{multline}
Thus, Sinkhorn drifting preserves the attraction--repulsion structure of the original drifting field in \eqref{eq:orignial_drifting_field}, but uses globally balanced transport weights instead of local kernel weights.
Please refer to \cite[Algorithm 1]{he2026sinkhorn} for the entire process to compute Sinkhorn drifting field for a mini-batch.
In the next subsection, we use this drifting field to train the enhancement model within the NyTT framework.

\subsection{NyTT with Sinkhorn drifting loss}

NyTT and its variants are typically based on signal-level reconstruction losses with respect to either the noisy speech in \cref{eq:nytt} or speech enhanced by a teacher model in \cref{eq:nyttsd}. 
These methods encourage the model to suppress the extra noise while preserving speech content, but the enhanced outputs may still retain observation noise or contain artifacts. 
We address this limitation by using the unpaired Sinkhorn drifting loss that encourages the distribution of enhanced speech to move toward the distribution of clean speech.

With the Sinkhorn drifting field from \cref{sec:sinkhorn_drifting}, we optimize the enhancement model $f_{\bm{\theta}}$ using the following loss function
\begin{equation}
    \mathcal{L}^{\mathrm{Drift}}(\bm{\theta})
    =
    \mathbb{E}_{\mathbf{x}, \mathbf{s}, \tilde{\mathbf n}}
    \left\|
    \hat{\mathbf z}
    -
    \texttt{sg}
    \left(
    \hat{\mathbf z}
    +
    \mathbf V_{\hat q_{\bm{\theta}},\hat p}^{\ell}(\hat{\mathbf z})
    \right)
    \right\|_2^2 ,
\label{eq:sinkhorn_loss}
\end{equation}
where $\tilde{\mathbf n}$ denotes additional noise used in NyTT, and 
\begin{equation}
    \hat{\mathbf z}
    =
    h
    \left(
    f_{\bm{\theta}}(\mathbf{x}+\tilde{\mathbf n})
    \right).
    \label{eq:encoder-proposed}
\end{equation}
We emphasize that the clean speech $\mathbf{s}$ in \eqref{eq:sinkhorn_loss} is sampled independently of $\mathbf{x}$, i.e., the proposed method does not necessarily require the pairs of $(\mathbf{x}, \mathbf{s})$.
Compared with the original loss in \cref{eq:original_loss} proposed in DriftSE~\cite{xu2026speech}, we first replace the mean-shift-based drifting field with the Sinkhorn drifting field $\mathbf V_{\hat q_{\bm{\theta}},\hat p}^{\ell}$.
Second, the vanilla Gaussian noise $\bm{\epsilon}$ in \eqref{eq:encoding} is replaced by the extra noise sampled from $p_\text{noise}$ following the formulation of NyTT.
Our final training objective combines the NyTT reconstruction loss with the mean-teacher framework in \cref{eq:nyttsd} and the Sinkhorn drifting loss in \cref{eq:sinkhorn_loss}:
\begin{equation}
    \mathcal{L}^{\mathrm{Proposed}}(\bm{\theta})
    =
    \mathcal{L}^{\mathrm{NyTT+}}(\bm{\theta})
    +
    \lambda
    \mathcal{L}^{\mathrm{Drift}}(\bm{\theta}),
    \label{eq:proposedloss}
\end{equation}
where $\lambda$ balances the contribution of the drifting loss.
While the first term enforces the sample-level faithfulness with respect to the observed signal $\mathbf{x}$, the second unpaired Sinkhorn drifting loss focuses on distribution-level matching between enhanced and clean speech.

To improve speech quality and naturalness, we further propose to use a speech quality assessment (SQA) model~\cite{wang2026urgentmos} as a speech encoder $h$ instead of self-supervised learning (SSL) models used in DriftSE~\cite{xu2026speech}.
In particular, we use UrgentMOS~\cite{wang2026urgentmos} as the SQA model and extract its utterance-level embedding before the final score-prediction head.
Since clean speech is expected to obtain high scores from the SQA model, by moving enhanced-speech features toward clean-speech features, the Sinkhorn drifting loss should also implicitly improve the score predicted by the SQA model.
The frame-level features from the SSL models convey phonetic information and provide high-resolution supervision in the CTT setting, where the underlying clean speech is in the same mini-batch.
However, matching phonetic information with unpaired speech might be unreasonable even though Sinkhorn drifting might find the optimal frame-level couplings.
We expect that using the features from an SQA model mitigates this issue.
In contrast to na\"{i}vely maximizing the score~\cite{wang2025asru}, our unpaired Sinkhorn drifting loss can take the diversity of clean speech into account.
We leave a detailed comparison  with the na\"ive maximization approach for future work.

\section{Experiments}

\subsection{Experiments on VoiceBank-DEMAND}
\label{sec:ex-vbdm}

\noindent\textbf{Datasets:}
We use a widely-used speech enhancement dataset, VoiceBank-DEMAND (VBDM)~\cite{valentinibotinhao2016}, to validate the efficacy of the proposed method under a controlled setting.
The 28-speaker subset is used for training, where either the first channel of noise recordings from the DEMAND dataset~\cite{thiemann2013demand} or artificially generated noise is added to clean utterances from the VoiceBank corpus~\cite{veaux2013}.
We reserve the speakers \texttt{p226} and \texttt{p287} for the validation set following~\cite{richter2023speech}.
All audio signals are downsampled to 16 kHz.

For NyTT, the additional noise $\tilde{\mathbf{n}}$ in \eqref{eq:nytt} is randomly sampled from the DEMAND dataset and the CHiME-3 dataset~\cite{Barker2015} following \cite{fujimura2021nytt}.
For DEMAND, we exclude all the channels of the recordings for environment types used in the test set and the first-channel recordings for all noise types to avoid leaking the exact same noise segment by using it as both the observation noise $\mathbf{n}$ and the additional noise $\tilde{\mathbf{n}}$.
The additional noise is rescaled so that SNR with respect to the observed signal $\mathbf{x}$ is distributed as $\mathcal{U}(-5, 5)$ in dB.
Since VBDM is a synthetic dataset, we can define a paired Sinkhorn drifting loss (P-Sinkhorn) using the underlying clean speech for $\mathbf{x}$ to compute the drift term.
Meanwhile, for the unpaired Sinkhorn drifting loss (UP-Sinkhorn), we use clean utterances from the EARS dataset~\cite{richter2024ears}, which ensures that the speaker and linguistic contents are mismatched with VBDM.
Note that we exclude non-speech sounds following the URGENT challenge (2025)~\cite{saijo2025urgent}.

\noindent\textbf{Models:}
We use a convolutional dual-path recurrent neural network (DPRNN) following recent weakly-supervised speech enhancement methods~\cite{jiang2026}.
The model takes real and imaginary parts and magnitude of the complex STFT coefficients as input and predicts a complex-valued time-frequency mask.
It consists of a three-layer Conv2D encoder, two time-frequency DPRNN blocks, and a three-layer ConvTrans2D decoder, together with a CBAM block~\cite{woo2018cbam}.
The model comprises around 1.5$\,$M parameters, excluding the EMA teacher model and the frozen SQA model for computing the drifting loss.
We use UrgentMOS~\cite{wang2026urgentmos} as the encoder $h$ in \eqref{eq:encoder-proposed}, where the feature dimension $D$ is 768.

\noindent\textbf{Training details:}
We compute STFT using a 510-point discrete Fourier transform with 128-sample window shift~\cite{richter2023speech}.
For NyTT with self-distillation in \cref{eq:nyttsd}, we use the standard negative SNR loss in the time domain.
For the Sinkhorn drifting, we average the drifting field computed with multiple $\tau$ in $(0.1, 0.5, 2.0)$ following ~\cite{deng2026generative,xu2026speech}.
The model is trained using the AdamW optimizer up to 200 epochs with a learning-rate warmup over 1000 steps up to $10^{-3}$.
During the training, each signal is segmented into two-second chunks, and the batch size is 32.
For EMA, we set the step-wise decay parameter to $0.8^{1/Q}$, where $Q$ is the number of training steps per epoch.
The Sinkhorn algorithm is iterated 10 times to compute the loss function, and the loss balancing parameter $\lambda$ in \eqref{eq:proposedloss} is set to 0.5.

\noindent\textbf{Evaluation setups:}
To evaluate the enhancement performance, we use two intrusive metrics, SI-SDR and wide-band PESQ.
We also report two non-intrusive metrics, DNSMOS P.808~\cite{reddy2021dnsmosp808} and UTMOS~\cite{saeki2022utmos}.

\begin{table}[t]
\centering
\caption{Enhancement performance on VoiceBank-DEMAND (VBDM), shown as mean scores with 95\% confidence intervals. The CTT rows are shown in orange, as it requires paired training data.
}
\label{tab:main}

\sisetup{
  reset-text-series=false,
  text-series-to-math=true,
  mode=text,
  minimum-decimal-digits=2,
  uncertainty-mode=separate,
  retain-zero-uncertainty=true,
  table-number-alignment=center,
  table-format=1.2(2)
}
\renewcommand{\arraystretch}{1.15}

\begin{threeparttable}
\resizebox{\linewidth}{!}{%
\begin{tabular}{lS[table-format=2.2(2)]*{3}{S[table-format=1.2(2)]}}
\toprule
Method & {SI-SDR} & {PESQ} & {DNSMOS} & {UTMOS} \\
\midrule
Noisy & \metric{8.444088733591167}{0.38415662102184467} & \metric{1.9700447381700126}{0.05148366053234014} & \metric{3.091904640197754}{0.02673033898494552} & \metric{3.1055857537730227}{0.052029560087548256} \\ 
\midrule
DriftSE~\cite{xu2026speech}           &  {\phantom{1}6.60} & {2.00} & {\bfseries 3.61} & {-} \\
unSE~\cite{jiang2023unse}               & {15.96} & {2.45} & {2.92} & {-} \\
RemixIt~\cite{tzinis2022remixit}$^\dagger$ & {17.68} & {\bfseries 2.47} & {-}  & {-} \\
SelfSE~\cite{jiang2026}$^\dagger$  & {\bfseries 18.07} & {2.40} & {-}  & {-} \\
\midrule
NyTT+ & 
  \bfseries \metric{18.584473230306383}{0.20967476135575003} &
  \metric{2.6330879672059737}{0.03899709760387637} & 
  \metric{3.5254452228546143}{0.019706982038200878} &
  \metric{3.810837360840399}{0.02396811941810985}
  \\
NyTT+ w/ UP-Drift & 
  \bfseries \metric{18.579660145981798}{0.20485107181382164} &
  \bfseries \metric{2.68579752034354}{0.04073481772441192} & 
  \bfseries \metric{3.5897200107574463}{0.0187578161122986} &
  \bfseries \metric{3.9318603754622266}{0.019998770378581145}
\\
\,\, no Sinkhorn & \metric{18.11365582642046}{0.20006047437620392} & \metric{2.61332646880335}{0.039922375495382856} & \metric{3.5377986431121826}{0.019159754311661326} & \metric{3.8200017830121866}{0.02468776915905923}\\
\,\, no SQA & \metric{18.550982672033957}{0.21033103551797616} & \metric{2.6335566994345303}{0.03974483277581475} & \metric{3.5289621353149414}{0.019763470521545843} & \metric{3.829208334962141}{0.024462418902428368} \\
\midrule
\rowcolor{ai2offwhite}
CTT &
  \bfseries \metric{19.737063678722937}{0.2513376582829177} &
  \metric{2.6963419976454337}{0.041904718204598096} &
  \metric{3.527475357055664}{0.02012467069135151} &
  \metric{3.7876187810330717}{0.022895118056609667}
  \\
\rowcolor{ai2offwhite}
CTT w/ P-Drift & \metric{19.711052302017954}{0.2517026995376295} & \bfseries \metric{2.774570611784759}{0.04405786760260402} & \bfseries \metric{3.552609920501709}{0.019848959144037906} & \bfseries \metric{3.893018316875384}{0.019630357541690183} \\
\rowcolor{ai2offwhite}
CTT w/ UP-Drift & \metric{19.707720496122118}{0.25025512407991046} & \metric{2.759742954691637}{0.04419958197712329} & \metric{3.544001579284668}{0.02019483479502927} & \metric{3.86260123464089}{0.0212200923549286} \\
\bottomrule
\end{tabular}%
}
\begin{tablenotes}[flushleft]
\footnotesize
\item[$\dagger$]
\parbox[t]{0.96\columnwidth}{
The scores are for the models trained with multi-resolution STFT
and SI-SNR losses, i.e., without a PESQ-specific loss, from the
SelfSE paper~\cite{jiang2026}.}
\end{tablenotes}
\end{threeparttable}
\end{table}

\noindent\textbf{Results:}
\Cref{tab:main} summarizes the enhancement performance for CTT and NyTT, together with existing unsupervised and weakly-supervised speech enhancement methods.
First, within CTT settings, colored in orange, incorporating the paired Sinkhorn drifting loss improves perceptually motivated metrics, i.e., PESQ, DNSMOS, and UTMOS, compared to using only the standard negative SNR loss.
The unpaired Sinkhorn drifting loss also improves these metrics while preserving a reasonable SI-SDR.
Hence, matching the enhanced-speech distribution with the clean-speech one is beneficial to improve perceptual metrics.

Next, our implementation of NyTT with a mean-teacher framework~\cite{song2026artt} (NyTT+) performs well compared with existing weakly-supervised and unsupervised methods.
The proposed combination, NyTT+ with the unpaired Sinkhorn drifting, further improves the perceptual metrics as in the case of CTT.
Compared with the unpaired drifting in the original DriftSE paper~\cite{xu2026speech}, the proposed combination substantially recovers intrusive metrics, e.g., SI-SDR from $6.60$ dB to $18.58$ dB, while retaining a similar DNSMOS score.
This result indicates that the unpaired drifting is effective for learning the clean-speech distribution while it is insufficient to ensure faithfulness with respect to a given observed signal.
Our proposed method mitigates this issue owing to the complementary nature of NyTT+.

As an ablation study, we investigate the performance of NyTT+ with the original drifting loss in the SQA model's latent space and with the Sinkhorn drifting loss in the DistillHuBERT's latent space.
When using DistillHuBERT, the Sinkhorn drifting loss is computed on each layer of DistillHuBERT and summed up as in \cite{xu2026speech}.
Both variants slightly improve UTMOS from NyTT+ without any drifting loss but substantially lag behind the proposed combination.
These ablations suggest that the combination of Sinkhorn drifting and the SQA latent representation contributes to the improvement.

\subsection{Experiments on CHiME-3 real recordings}

\noindent\textbf{Setup:}
To validate the efficacy of the proposed method on real recordings, we perform a domain adaptation experiment on the CHiME-3 real recordings~\cite{Barker2015}.
We follow the official data split, but exclude the 7138 simulated utterances from the training set.
The fifth channel is always used following \cite{Zhang2023}.
We fine-tune the enhancement model pre-trained on VBDM on the real recordings using NyTT variants, including ours.
For a fair comparison, we use the model pre-trained only with the negative SNR loss for initialization (CTT in Table~\ref{tab:main}).
The fine-tuning is performed up to 100 epochs using the AdamW optimizer as in Section~\ref{sec:ex-vbdm} but without learning-rate warmup.
As in the previous experiment, the additional noise is sampled from the DEMAND dataset and the CHiME-3 dataset, and unpaired clean speech is taken from the EARS dataset.
The proposed loss function is computed with multiple $\tau$ in $(0.1, 0.5, 2.0)$ and $\lambda$ is set to $0.5$.
Since the underlying clean speech is not available for the real recordings, we use the word error rates (WERs) and SI-SDR estimated by Torchaudio-Squim~\cite{Kumar2023} in addition to the perceptual non-intrusive metrics.
The Whisper Large v2 model is used for recognition following \cite{Zhang2023}.

\noindent\textbf{Results:}
\Cref{tab:chime3} presents the results on real recordings from the CHiME-3 evaluation set.
Compared with the CTT model trained on out-of-domain data (VBDM), the models fine-tuned with NyTT substantially reduce WER.
The model fine-tuned only using NyTT with the mean-teacher framework in \cref{eq:nyttsd} does not improve DNSMOS.
On the other hand, incorporating the proposed unpaired Sinkhorn drifting loss leads to better DNSMOS and UTMOS.
We confirm the statistical significance of these improvements over NyTT+ under paired $t$-tests ($p < 10^{-10}$).
Unfortunately, the proposed method still results in higher WER compared to the unprocessed noisy signal.
This may be because Whisper was trained to be robust against natural background noise but might not have been exposed to enhancement artifacts~\cite{iwamoto22interspeech}.

\begin{table}[t]
\caption{DNSMOS, UTMOS, and recognition performance on real recordings from the CHiME-3 evaluation set.}
\label{tab:chime3}
\sisetup{
    reset-text-series = false, 
    text-series-to-math = true, 
    mode=text,
    tight-spacing=true,
    round-mode=places,
    uncertainty-mode=separate,
    retain-zero-uncertainty=true,
    round-precision=2,
    table-format=2.2(2),
    table-number-alignment=center
}
\resizebox{\linewidth}{!}{
\begin{tabular}{lS[round-mode=places, round-precision=1, table-format=2.1]*{3}{S}}
\toprule
Method
& {WER (\%)}
& {SI-SDR*}
& {DNSMOS}
& {UTMOS}
\\
\midrule
Noisy & \bfseries  6.7 & \metric{0.8666209460913458}{0.23373544877199226} & \metric{2.467233180999756}{0.011692132997077354} & \metric{1.3857021629810333}{0.01278333978305538} \\
\midrule
CTT$^\dagger$ & 16.276 & \metric{8.7401155143694}{0.2617794521407157} & \metric{3.4254283905029297}{0.017957072444966728} & \metric{2.5385510461800025}{0.0339256799789639} \\
NyTT+ & 9.281 & \metric{9.389523559131405}{0.1972304315020292} & \metric{3.418638229370117}{0.016126169376407172} & \metric{2.596088872985406}{0.030166313423742062} \\
NyTT+ w/ UP-Drift & 9.115 & \bfseries \metric{10.159911370362071}{0.1866095461655201} & \bfseries \metric{3.5881550312042236}{0.014949076347786307} & \bfseries \metric{2.81728378507224}{0.03000639995723331} \\
\bottomrule
\end{tabular}
}
\begin{tablenotes}[flushleft]\footnotesize
\item[*] $^*$SI-SDR is estimated by Torchaudio-Squim~\cite{Kumar2023}.
\item[*] $^\dagger$CTT is performed on the VBDM dataset, i.e., there is a domain mismatch.
\end{tablenotes}
\end{table}

\section{Conclusion}

In this paper, we propose to combine NyTT with unpaired Sinkhorn drifting for weakly supervised speech enhancement.
By leveraging unpaired clean speech, we aim to align the enhanced-speech distribution with the clean-speech distribution in the latent space of an SQA model.
In our experiments on VBDM, the proposed combination improves several perceptual metrics over NyTT with the mean-teacher framework.
We also demonstrate the effectiveness of our method in a domain adaptation experiment on the real recordings from CHiME-3.

\bibliographystyle{IEEEtran}
\bibliography{strings,refs}

\end{document}